\documentclass[11pt]{article}
\usepackage[]{graphicx}
\usepackage{amsmath,amssymb}
\newcommand{\micron}{\mu\mbox{m}}
\begin{document}
\title{Efficient optical path folding by using multiple total
  internal reflections in a microcavity}

\author{Susumu Shinohara,$^1$ Satoshi Sunada,$^2$ Takehiro
  Fukushima,$^3$\\ Takahisa Harayama,$^4$ Kenichi Arai,$^1$ and
  Kazuyuki Yoshimura$^1$\\ \\
$^1$ NTT Communication Science Laboratories,\\ NTT Corporation, Kyoto
619-0237, Japan\\
$^2$ Faculty of Mechanical Engineering, Institute of Science
and\\ Engineering, Kanazawa University, Ishikawa 920-1192, Japan\\
$^3$ Department of Information and Communication
Engineering,\\ Okayama Prefectural University, Okayama 719-1197,
Japan\\
$^4$ Department of Applied Physics, School of Advanced Science\\ and
Engineering, Waseda University, Tokyo 169-8555, Japan }
\maketitle
\begin{abstract}
We propose using an asymmetric resonant microcavity for the efficient
generation of an optical path that is much longer than the diameter of
the cavity. The path is formed along a star polygonal periodic orbit
within the cavity, which is stable and confined by total internal
reflection. We fabricated a semiconductor device based on this idea
with an average diameter of 0.3 mm, and achieved a path length of 2.79
mm experimentally.
\end{abstract}
\vskip 10mm
Asymmetric resonant cavities (ARCs) such as deformed disks and
deformed spheres were proposed to make it possible to extract
directional emissions while maintaining relatively high quality
factors \cite{Noeckel97}.
The effects of deformations on resonant mode characteristics have been
studied by using the concepts and techniques of the quantum/wave chaos
theory, and the idea of using cavity shape as a design parameter for
controlling emission characteristics \cite{Schwefel05} has now been
well established \cite{Harayama11a}, with a representative achievement
being a uni-directional emission deformed microdisk \cite{Wiersig11}.

An interesting but less explored feature of ARCs is that they can
generate long optical paths by using multiple reflections at cavity
interfaces.
This feature was demonstrated for a macroscopic (i.e., cm-size)
deformed sphere made of copper with the aim of using it for gas
sensing \cite{Narimanov07, Qu08a, Qu08b}, where a long optical path is
important for increasing the sensitivity.
In that work, it was reported that a three-dimensional ARC with a
diameter of 5.24 cm can achieve a path length up to 6.0 m.
The idea itself of generating a long optical path by multiple
reflections can be found in traditional gas sensing cavities
consisting of multiple mirrors such as White cells \cite{White42} and
Herriott cells \cite{Herriott65}, where the careful alignment of
mirrors is necessary for generating a desired optical path.
An advantage of using ARCs for long path generation is that they are
alignment-free and suitable for miniaturization.

In addition to gas sensing, there are many applications where long
optical paths are necessary or desirable, e.g. optical delay lines
\cite{Lee12}, optical gyroscopes \cite{Lefevre93}, laser frequency
stabilization \cite{Lee13}, and laser chaos generation
\cite{Soriano13}.
If there is no device size limitation, optical fibers would provide
solutions in most cases.
However, for micro-scale devices where the use of optical fibers is
irrelevant, the idea of ARCs appears to be a promising option.
For instance, recently, the chaotic output from a semiconductor laser
with delayed optical feedback has been utilized for Gbps-speed
physical random bit generation \cite{Uchida08}, and efforts have been
made to reduce the device size by photonic integration
\cite{Argyris08, Harayama11b, Takahashi14}.
To ensure good quality of randomness, an external cavity at least
about 3 mm long is indispensable for delay generation
\cite{Takahashi14}, which currently is an obstacle for further
miniaturization.
The use of ARCs has the potential to provide a breakthrough as regards
this problem \cite{Sunada14}.

In this Letter, we propose the use of a deformed annular microdisk for
efficiently achieving a path length much longer than the device's
diameter, while suppressing refraction losses by total internal
reflection (TIR).
We fabricate a semiconductor device based on this idea, and show that
a path length of 2.79 mm can be achieved by the device, whose average
diameter is 0.3 mm.
In previous studies~\cite{Narimanov07, Qu08a, Qu08b}, a long path was
attributed to a multi-reflected chaotic ray orbit.
However, it is difficult to identify the responsible orbit and thus to
predict {\it a priori} an optical path length (or equivalently a delay
time).
In contrast, here we systematically construct a stable periodic ray
orbit within the cavity to obtain a long optical path, which enables
us to predict the path length precisely.
The work in Ref.~\cite{Sunada14} uses a two-dimensional
microcavity with quasi-stadium shape to form such a stable periodic
ray orbit, and light propagation along it was confirmed.
However, most of the light intensity is lost through transmission,
since the incident angles at the cavity interfaces are close to 0
degrees.
A significant improvement demonstrated in the current work is that a
stable ray orbit is reflected more times by a cavity interface, while
all of the reflections are TIRs, so that the confinement is more
efficient in terms of energy and space.

\begin{figure}
\begin{center}
\includegraphics[width=0.7\textwidth]{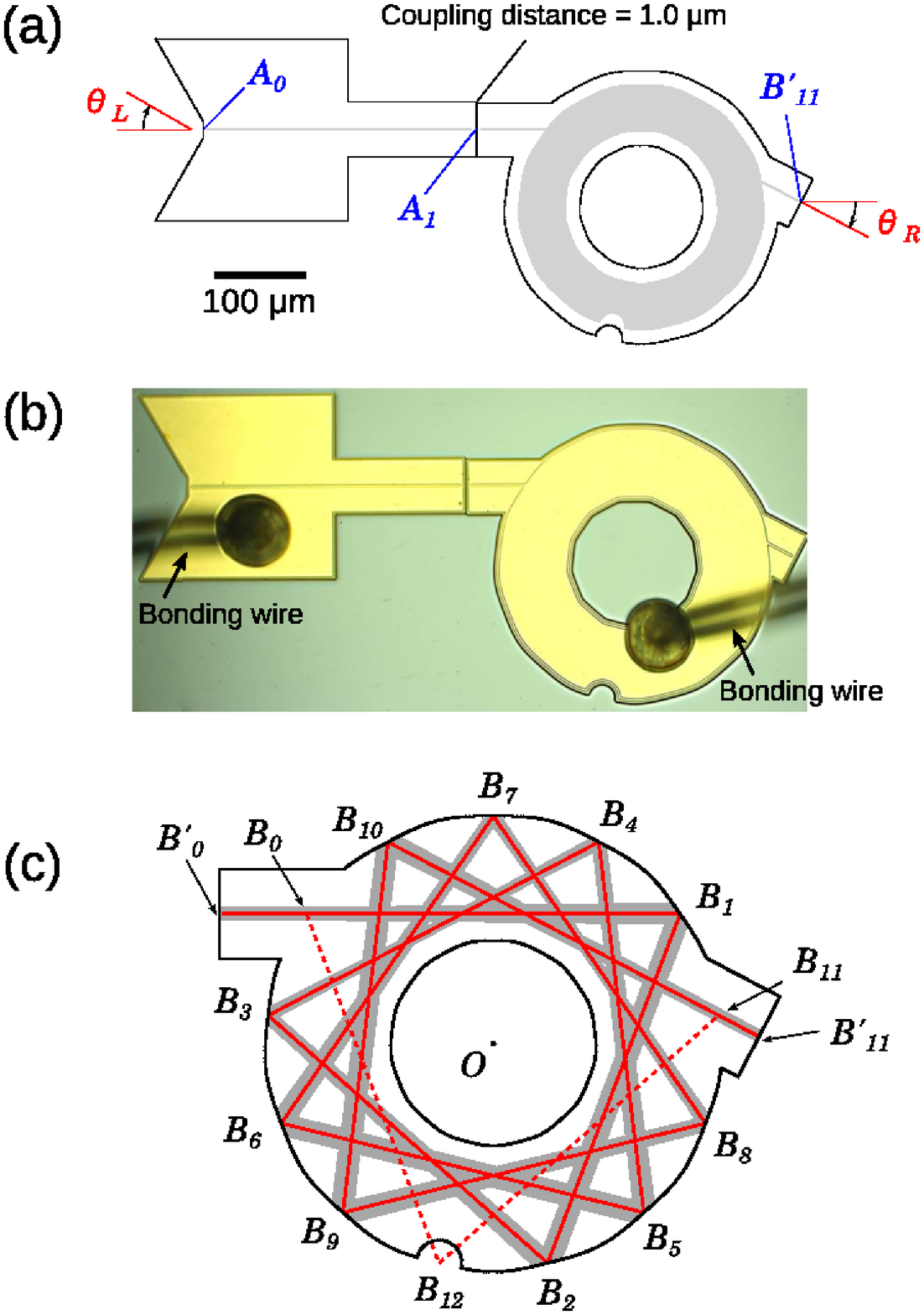}
\end{center}
\caption{\label{fig:cavity} (a) The geometry of a deformed annular
  cavity (right) coupled to a quasi-stadium cavity (left). In the
  deformed annular cavity, a path length of 2.79 mm is formed (see
  text). The light is emitted from the left facet $A_0$ and the right
  facet $B'_{11}$. The pumping regions are shown in gray. (b) Optical
  microscope image of the fabricated device. Bonding wires are
  attached on top of both cavities. (c) The deformed annular cavity
  with a self-retracing star polygonal orbit connecting the two facets
  $B_0'$ and $B_{11}'$. The orbit is confined by total internal
  reflection except for the bounces at $B_0'$ and $B_{11}'$. The beam
  width estimated from the Gaussian-optical theory is indicated by the
  gray area along the star polygonal orbit. The facets $B_0'$ and
  $B_{11}'$ correspond to the beam waist positions. The beam waist
  size at $B_0'$ is $w_{B_0'}=4.37\,\micron$, while the beam waist
  size at $B'_{11}$ is $w_{B'_{11}}=4.08\,\micron$.}
\end{figure}

Figure \ref{fig:cavity} shows the geometry of our device, which
consists of a quasi-stadium cavity (left) and a deformed annular
cavity (right).
Both cavities are pumped, but in this Letter, we consider the former
to be a light source and the latter to be an external cavity (i.e.,
the pumping for the latter is set below the threshold).
The quasi-stadium cavity has Fabry-Perot-type modes
\cite{Fukushima04}, while the deformed annular cavity has modes whose
intensities localize along a ray orbit with a long path reflected by
TIR (described below).
With this design, the light emitted from the quasi-stadium cavity is
coupled to the deformed annular cavity, amplified there, and then
emitted.
The light is emitted to the far field from the left facet $A_0$ of the
quasi-stadium cavity and from the output port facet $B_{11}'$ of the
deformed annular cavity [see Fig. \ref{fig:cavity} (a)].

Here, we define the cavities more precisely:
With the quasi-stadium cavity, modes are formed along the
Fabry-Perot-type orbit connecting both ends $A_0$ and $A_1$, whose
distance is $L_{FP}=300\,\micron$.
The radii of curvature of the cavity boundaries at $A_0$ and $A_1$ are
600\,$\micron$ and $\infty$, respectively.
These curvature values make the orbit stable \cite{Siegmann86}, and
corresponding resonant modes are given by Gaussian beams whose beam
waist is at $A_1$ with a (half) width of $w_{A_1}=5.00\,\micron$.

The boundary of the deformed annular cavity is defined in the polar
coordinates $(r,\phi)$ by $r(\phi)=R(1-\epsilon \sin(Q \phi))$, where
$R=r_0$ for the outer boundary, while $R=r_1$ for the inner boundary.
Here we fix the ratio as $r_1/r_0=0.45$.
When we consider this cavity as a closed billiard, it has a star
polygonal periodic orbit with $Q$ vertices.
We use $P$ to denote the number of clockwise rotations around the
origin $O$ for the ray propagation along the star polygonal orbit
until it returns to the initial point (an example for $Q=13$ and $P=4$
is shown in Fig. \ref{fig:cavity} (c)).

When setting the $Q$ and $P$ values, we consider the two factors
described below.
(i) The total length $L_{\star}$ of the star polygonal orbit is given by
\begin{equation}
{\displaystyle L_{\star}=2 r_0 \left(1-\epsilon\right) Q \sin(\pi P/Q)}.
\end{equation}
As we are interested in generating as long a path as possible, it is
desirable to set $P/Q$ close to $1/2$ with as large a $Q$ value as
possible.
Here, we also have to note that the distance between adjacent vertices
of the star polygonal orbit (e.g., the arc $B_4 B_1 \approx 2\pi
r_0/Q$) needs to be much longer than a wavelength within the cavity,
which limits the maximum value of $Q$.
This is a condition imposed by the wave picture to guarantee that the
wave propagates distinctly along the orbit.
(ii) The incident angle at the cavity boundary is given by
$\theta_{in}=\pi(Q-2P)/(2Q)$.
Since we want to confine the orbit by TIR, we have the condition
$\theta_{in}>\sin^{-1}(1/n_{eff})$, where $n_{eff}$ is the effective
refractive index of the cavity, and we assume the cavity is surrounded
by air.
In this work, we fix $r_0=150\,\micron$, and assume that the cavity is
made of GaAs (i.e., the wavelength in vacuum $\lambda\approx
0.86\,\micron$ and $n_{eff}=3.3$).
To satisfy the above two conditions sufficiently, we choose $Q=13$ and
$P=4$ in the current work.

Next, we explain how we determine the value of the parameter
$\epsilon$.
We open the deformed annular cavity so that it couples to the
quasi-stadium cavity and to the far field.
We attach an input port with length $d_{in}=\overline{B_0 B_0'}$ at
the vertex $B_0$ and an output port with length
$d_{out}=\overline{B_{11}B_{11}'}$ at the vertex $B_{11}$ as shown in
Fig. \ref{fig:cavity} (c).
Then, we have a self-retracing periodic orbit connecting the facets of
the input and output ports, $B'_0$ and $B'_{11}$.
To allow us to assume that the light propagates along this orbit in
the form of Gaussian beams, we make the orbit linearly stable
\cite{Tureci02}.
This condition is expressed by $|\mbox{Tr}M|<2$ with $M$ being the
monodromy matrix for the periodic orbit.
To fulfill this condition, we chose $\epsilon=0.005$,
$d_{in}=57.17\,\micron$ and $d_{out}=30.0\,\micron$, which yield
$|\mbox{Tr}M|=0.12$.
This trace value, together with the monodromy matrix elements, is
directly related to the beam width of the Gaussian beam
\cite{Tureci02}.
Denoting the coordinate along the star polygonal orbit as $z$, and the
coordinate perpendicular to it as $x$, we can express the Gaussian
beam profile as $\exp[-x^2/w^2(z)]$, where $w(z)$ represents the beam
width.
In Fig. \ref{fig:cavity} (c), the area corresponding to the beam width
(i.e., $|x|\leq w(z)$) is shown in gray.
The input port facet $B'_0$ corresponds to the beam waist position,
where the beam width is $w_{B'_0}=4.37\,\micron$.
This matches the beam waist size of the quasi-stadium cavity at the
facet $A_1$, $w_{A_1}=5.00\,\micron$.
The coupling distance $\overline{A_1 B'_0}$ is set at $1.0\,\micron$.
We also note that a half-circular scatterer with a radius $0.1\times
r_0=15\,\micron$ is placed at the vertex $B_{12}$ to suppress the
existence of undesired high-Q modes.

With the parameter values specified above, the total (one-way) length
of the TIR-confined orbit is $L_{tot}=2.79$\,mm, which is 9.3 times
longer than the diameter of the deformed annular cavity.
Below, we demonstrate the existence of this long path experimentally.
We fabricated a device with the cavity design described above by
applying a reactive-ion-etching technique to a graded index separate
confinement heterostructure single quantum well GaAs/AlGaAs structure
grown by metal organic chemical vapor deposition.
The layer structure and fabrication process are described in detail in
Ref. \cite{Fukushima04}.
For the quasi-stadium cavity, the electrode contact is formed only
along the Fabry-Perot orbit with a narrow width of $2\,\micron$, while
for the deformed annular cavity, the electrode contact is patterned to
amplify the star polygonal orbit (the gray regions in
Fig. \ref{fig:cavity} (a) represent the pumping regions).
For the measurements, the device was operated in the continuous-wave
mode at 25$^{\circ}$C.
In what follows, we denote the injection current supplied to the
quasi-stadium cavity as $I_{LD}$, and that supplied to the deformed
annular cavity as $I_{SOA}$, to indicate that the former cavity is a
laser diode (LD), while the latter works as a semiconductor optical
amplifier (SOA).
The lasing threshold for the quasi-stadium cavity was
$I_{LD}=60\,\mbox{mA}$, while that for the deformed annular cavity was
$I_{SOA}=124\,\mbox{mA}$.
These thresholds were almost unchanged when the other cavity was
pumped.

\begin{figure}
\begin{center}
\includegraphics[width=0.7\textwidth]{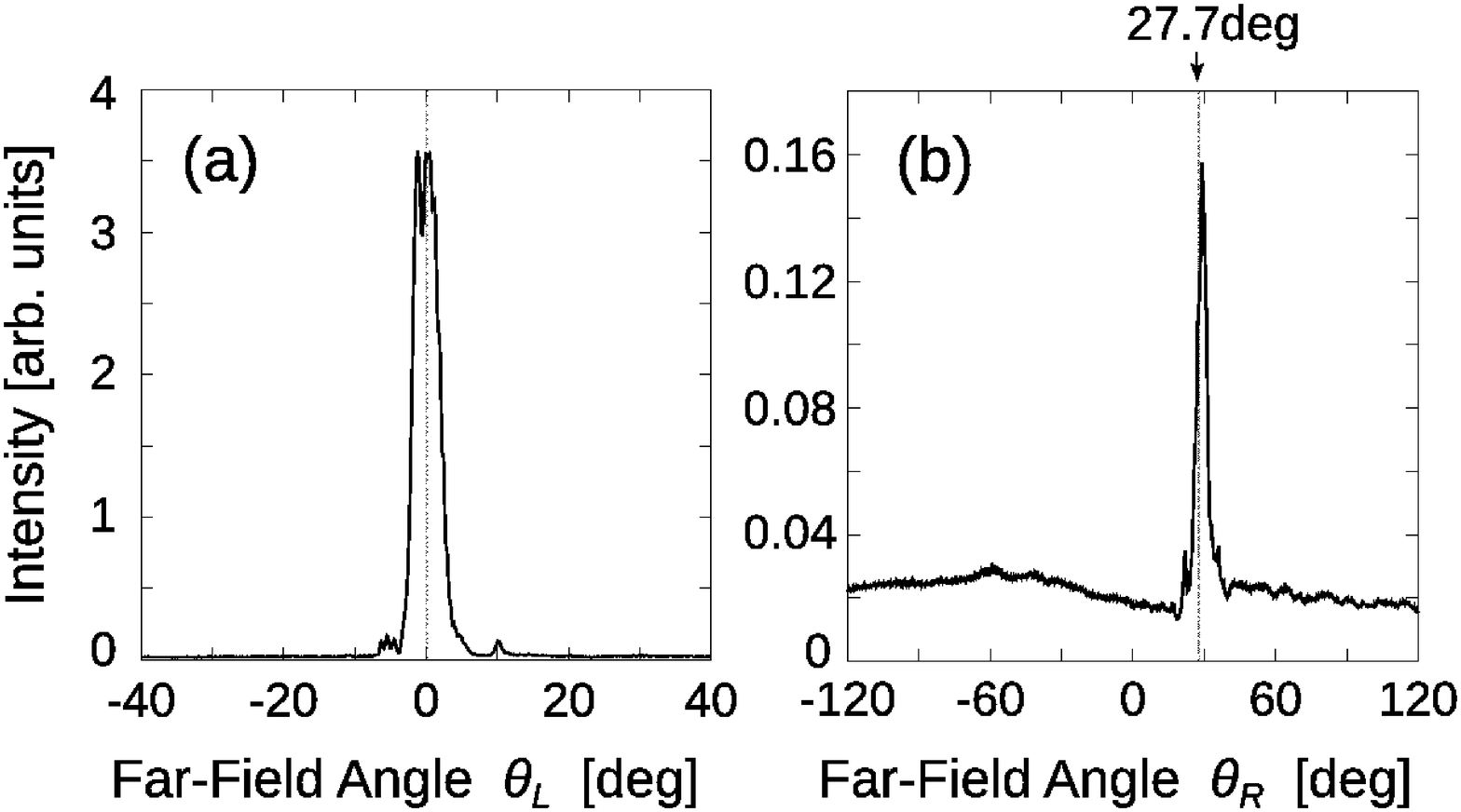}
\end{center}
\caption{\label{fig:ffps} Measured far-field emission patterns: (a)
  Output from facet $A_0$ of the quasi-stadium cavity. (b) Output
  from facet $B'_{11}$ of the deformed annular cavity. The angles
  $\theta_L$ and $\theta_R$ are defined in Fig. \ref{fig:cavity} (a).}
%\end{figure}
%
%\begin{figure}
\vskip 10mm
\begin{center}
\includegraphics[width=0.5\textwidth]{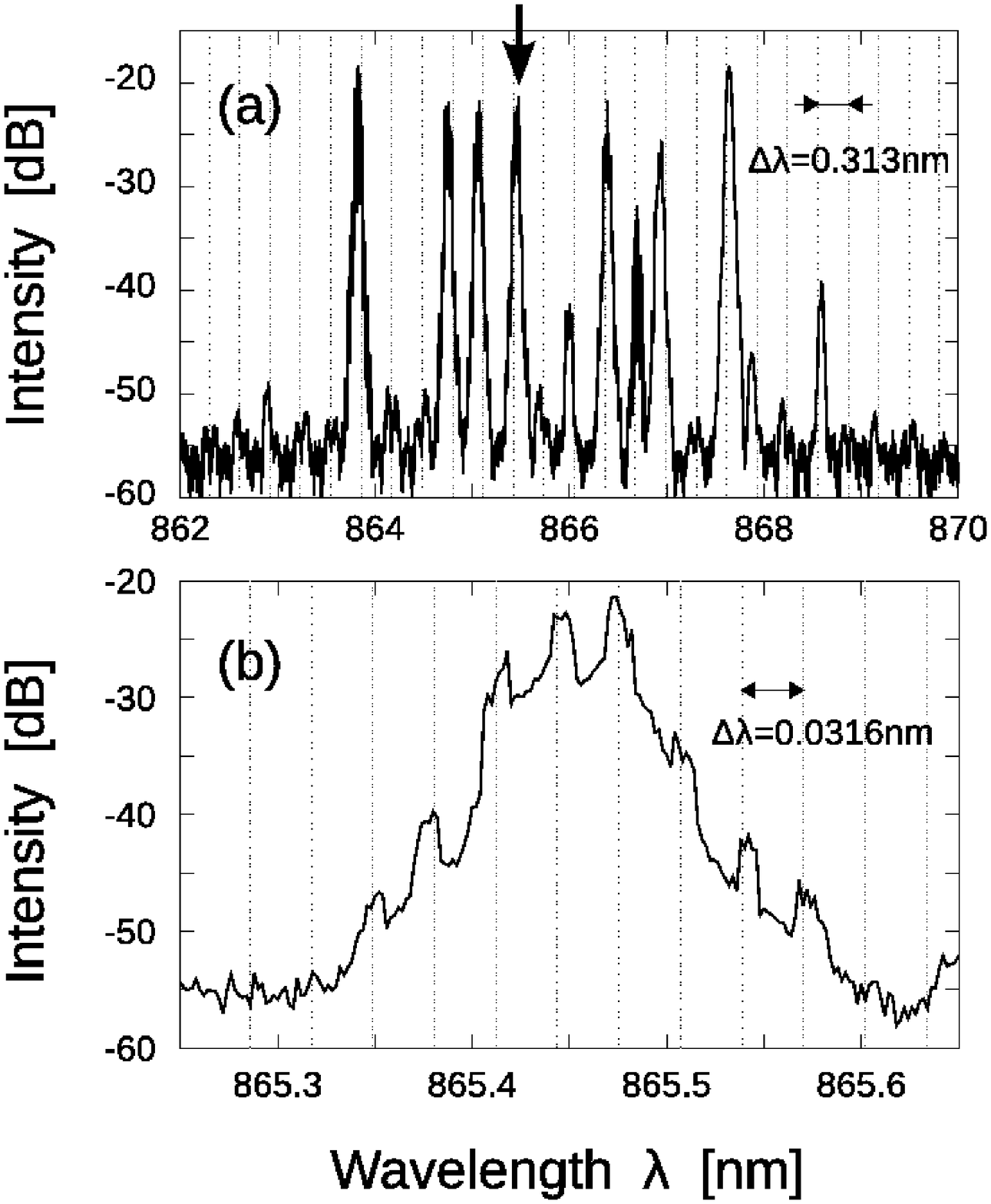}
\end{center}
\caption{\label{fig:opt-spectra}
Optical spectra for the output from facet $A_0$ of the quasi-stadium
cavity for $I_{LD}$=$150\,\mbox{mA}$ and $I_{SOA}$=$120\,\mbox{mA}$:
(a) shows the whole range, while (b) shows an enlarged image of the
peak at $\lambda\approx 865.45$ nm (indicated by an arrow in
Fig. \ref{fig:opt-spectra} (a)).
Regular modal spacings $\Delta\lambda=0.313\,\mbox{nm}$ and
$\Delta\lambda=0.0316\,\mbox{nm}$ are indicated by dashed vertical
lines in (a) and (b), respectively (see text for details).
}
\end{figure}

Figure \ref{fig:ffps} (a) and (b) show the measured far-field emission
patterns obtained when $I_{LD}=120\,\mbox{mA}$ (far above the
threshold) and $I_{SOA}=120\,\mbox{mA}$ (just below the threshold) for
the output from the quasi-stadium cavity [Fig. \ref{fig:ffps} (a)] and
for the output from the deformed annular cavity [Fig. \ref{fig:ffps}
  (b)].
In Figs. \ref{fig:ffps} (a) and \ref{fig:ffps} (b), we can confirm the
outputs from the quasi-stadium cavity and the deformed annular cavity,
respectively.
The position of the sharp peak in Fig. \ref{fig:ffps} (b),
$\theta_R=27.7$ degrees is in good agreement with the angular
direction of the output port of the deformed annular cavity.
The far-field peak at around $\theta_R=27.7$ degrees was clearly
observed when $I_{LD}$ and $I_{SOA}$ were sufficiently large (for
example, for $I_{SOA}=120\,\mbox{mA}$, we could observe a sharp peak
for $I_{LD}\gtrsim 80\,\mbox{mA}$).
Assuming that the far-field intensity is expressed as $I\propto
\exp[-2 (\theta-\theta_0)^2/\sigma^2]$, we can evaluate the beam
divergence as $\sigma=3.98\,\mbox{degrees}$ in Fig. \ref{fig:ffps} (b).
This value is in good consistence with the beam waist size of the
output beam at $B'_{11}$, $w_{B'_{11}}=4.08\,\micron$, which yields
$\sigma=\lambda/(\pi w_{B'_{11}})=3.87\,\mbox{degrees}$.
These results validate our assumption that the light propagates along
the star polygonal orbit.
As is observed in Figs. \ref{fig:ffps} (a) and (b), the output power
from the facet $B_{11}'$ is reduced by an order of magnitude as
compared with the output power from the facet $A_0$.
We expect that this is caused by coupling losses at an air gap
between the two facets $A_1$ and $B_0'$, and gain saturation in the
deformed annular cavity.

To further confirm the light propagation along the star polygonal
orbit as well as the coupling of the two cavities, we measured the
optical spectra for the emission from the quasi-stadium cavity (i.e.,
output from facet $A_0$).
Figure \ref{fig:opt-spectra} shows data for $I_{LD}=150\,\mbox{mA}$
and $I_{SOA}=120\,\mbox{mA}$, where Fig. \ref{fig:opt-spectra} (a) is
for the whole range, while Fig. \ref{fig:opt-spectra} (b) is an
enlarged image of the peak at $\lambda\approx 865.45$ nm in (a)
(indicated by an arrow).
In Fig. \ref{fig:opt-spectra} (a), we can see a regular modal spacing
of $\Delta\lambda=0.313\,\mbox{nm}$.
This value corresponds well to the length of the quasi-stadium cavity,
$L_{FP}=300\,\mu m$, which yields a longitudinal modal spacing of
$\Delta\lambda=0.30\,\mbox{nm}$ evaluated from the formula
$\Delta\lambda=\lambda^2/
\{2n_{eff}L_{FP}[1-(\lambda/n_{eff})(dn_{eff}/d\lambda)]\}$ with the
dispersion being $dn_{eff}/d\lambda=-1\,\micron^{-1}$ (taken from
Ref. \cite{Casey78}).
Meanwhile, in Fig. \ref{fig:opt-spectra} (b), we can observe a regular
modal spacing of $\Delta\lambda=0.0316\,\mbox{nm}$.
For each peak in Fig. \ref{fig:opt-spectra} (a), we could observe a
similar substructure with the same modal spacing.
This modal spacing value is explained by the path length of the
coupled cavities, $L_{FP}+L_{tot}=3.09$\,mm, which yields
$\Delta\lambda=0.029\,\mbox{nm}$.
These results mean that the characteristics of both quasi-stadium
cavity and deformed annular cavity are manifested in the output of the
quasi-stadium cavity laser.
This validates our assumption regarding the coupling of the two
cavities as well as the propagation of light along the star polygonal
orbit.

In conclusion, we proposed the use of an asymmetric resonant
microcavity for systematically constructing a long path corresponding
to a stable star polygonal orbit that is confined by TIRs except for
the input and output ports.
We illustrated this idea by fabricating a device designed to generate
a path length of 2.79 mm, which is 9.3 times longer than the cavity's
diameter.
In principle, this factor can be greatly increased by optimizing the
values of $Q$ and $P$.
From the measured far-field emission patterns and optical spectra, we
confirmed for the fabricated device that the light propagates along
the designed long path.
As reported in Ref. \cite{Takahashi14}, a 3 mm-long external
cavity is sufficient for generating chaotic oscillations suited for
random bit generation.
Since the extension of the path length over 3 mm is well within reach
for our cavity design, we think that it can be used for realizing a
compact random bit generator, where it is preferable to fabricate the
cavity by a passive material and to apply high-reflection coating at
the facet $B_{11}'$.
We also expect that our idea would be of use for realizing a compact
information processing device based on optical delay feedback
\cite{Soriano13}.
As for the scalability, it appears to be of interest to couple
multiple cavities to generate a much longer path while densely filling
a two-dimensional area with them.
This would allow us to make a path length proportional to $l^2$ where
$l$ is the linear dimension of the footprint.
These ideas will be examined in future works.
\vskip 10mm
S. Sunada was supported in part by a Grant-in-Aid for Young Scientists
(B) (Grant No.26790056) from the MEXT of Japan.
%

%
%\bibliographystyle{unsrt}
%\bibliography{ssttkk}
%
%\newpage
\newpage
\end{document}